\documentclass[aps,twocolumn,pra,tightenlines,floatfix,superscriptaddress]{revtex4-2}

\usepackage[breaklinks=true]{hyperref}
\usepackage{graphicx}
\usepackage[english]{babel}
\usepackage{amsmath}
\usepackage{amssymb}
\usepackage{times}
\usepackage{lipsum}

\begin{document}

\title{Effects of  particle-hole fluctuations on the  superfluid transition in two-dimensional atomic Fermi gases}

 \author{Junru Wu}
 \affiliation{Hefei National Research Center for Physical Sciences at the Microscale and School of Physical Sciences, University 
   of Science and Technology of China, Hefei, Anhui 230026, China}
 \affiliation{Shanghai Research Center for Quantum Science and CAS Center for Excellence in Quantum Information and Quantum Physics, 
 University of Science and Technology of China, Shanghai 201315, China}
 \affiliation{Hefei National Laboratory, University of
  Science and Technology of China, Hefei 230088, China}
 \author{Zongpu Wang}
 \affiliation{Hefei National Research Center for Physical Sciences at the Microscale and School of Physical Sciences, University 
   of Science and Technology of China, Hefei, Anhui 230026, China}
 \affiliation{Department of Physics, University of Hong Kong, Pokfulam Road, Hong Kong}
 \author{Lin Sun}
 \affiliation{Hefei National Laboratory, University of
  Science and Technology of China, Hefei 230088, China}
 \affiliation{Shanghai Research Center for Quantum Science and CAS Center for Excellence in Quantum Information and Quantum Physics, 
 University of Science and Technology of China, Shanghai 201315, China}
 \author{Kaichao Zhang}
 \affiliation{Hefei National Research Center for Physical Sciences at the Microscale and School of Physical Sciences, University 
   of Science and Technology of China, Hefei, Anhui 230026, China}
 \affiliation{Shanghai Research Center for Quantum Science and CAS Center for Excellence in Quantum Information and Quantum Physics, 
 University of Science and Technology of China, Shanghai 201315, China}
 \affiliation{Hefei National Laboratory, University of
  Science and Technology of China, Hefei 230088, China}
 \author{Chuping Li}
 \affiliation{Hefei National Research Center for Physical Sciences at the Microscale and School of Physical Sciences, University 
   of Science and Technology of China, Hefei, Anhui 230026, China}
 \affiliation{Shanghai Research Center for Quantum Science and CAS Center for Excellence in Quantum Information and Quantum Physics, 
 University of Science and Technology of China, Shanghai 201315, China}
 \affiliation{Hefei National Laboratory, University of
  Science and Technology of China, Hefei 230088, China}
 \author{Yuxuan Wu}
 \affiliation{Hefei National Research Center for Physical Sciences at the Microscale and School of Physical Sciences, University 
   of Science and Technology of China, Hefei, Anhui 230026, China}
 \affiliation{Shanghai Research Center for Quantum Science and CAS Center for Excellence in Quantum Information and Quantum Physics, 
 University of Science and Technology of China, Shanghai 201315, China}
 \affiliation{Hefei National Laboratory, University of
  Science and Technology of China, Hefei 230088, China}
 \author{Pengyi Chen}
 \affiliation{Hefei National Research Center for Physical Sciences at the Microscale and School of Physical Sciences, University 
   of Science and Technology of China, Hefei, Anhui 230026, China}
 \affiliation{Shanghai Research Center for Quantum Science and CAS Center for Excellence in Quantum Information and Quantum Physics, 
 University of Science and Technology of China, Shanghai 201315, China}
 \affiliation{Hefei National Laboratory, University of
  Science and Technology of China, Hefei 230088, China}
 \author{Dingli Yuan}
 \affiliation{Hefei National Research Center for Physical Sciences at the Microscale and School of Physical Sciences, University 
   of Science and Technology of China, Hefei, Anhui 230026, China}
 \affiliation{Shanghai Research Center for Quantum Science and CAS Center for Excellence in Quantum Information and Quantum Physics, 
 University of Science and Technology of China, Shanghai 201315, China}
 \affiliation{Hefei National Laboratory, University of
  Science and Technology of China, Hefei 230088, China}
 \author{Qijin Chen}
 \email[Corresponding author: ]{qjc@ustc.edu.cn}
 \affiliation{Hefei National Research Center for Physical Sciences at the Microscale and School of Physical Sciences, University 
   of Science and Technology of China, Hefei, Anhui 230026, China}
 \affiliation{Shanghai Research Center for Quantum Science and CAS Center for Excellence in Quantum Information and Quantum Physics, 
 University of Science and Technology of China, Shanghai 201315, China}
 \affiliation{Hefei National Laboratory, University of
  Science and Technology of China, Hefei 230088, China}

\date{\today}

\begin{abstract}
  Proper treatment of the many-body interactions is of paramount
  importance in our understanding of strongly correlated systems.
  Here we investigate the effects of particle-hole fluctuations on the
  Berezinskii-Kosterlitz-Thouless (BKT) transition in two-dimensional
  Fermi gases throughout the entire BCS-BEC crossover.  We include
  self-consistently in the self energy treatment the entire
  particle-hole $T$ matrix, which constitutes a renormalization of the
  bare interaction that appears in the particle-particle scattering
  $T$ matrix, leading to a screening of the pairing interaction and
  hence a reduction of the pairing gap and the transition temperature
  $T_\text{BKT}$.  Here $T_\text{BKT}$ is determined by the critical
  phase space density, for which the pair density and pair mass are
  determined using a pairing fluctuation theory, which accommodates
  self-consistently the important self-energy feedback in the
  treatment of finite-momentum pairing fluctuations.  The screening
  strength varies continuously from its maximum in the BCS limit to
  essentially zero in BEC limit. In the unitary regime, it leads to an
  interaction-dependent shift of $T_\text{BKT}$ towards the BEC
  regime. This shift is crucial in explaining experimental data, when
  quantitativeness is important, as they often depend on the
  interaction strength.  Where experimental and simulation data are
  available, our findings are consistent with experiment in the
  unitary and BEC regimes and with quantum Monte Carlo simulations in
  the BCS and unitary regimes.
\end{abstract}

\maketitle

\section{Introduction}

Strongly correlated systems constitute a main challenge and are thus
at the heart and frontier of condensed matter physics. Due to the
low dimensionality, fluctuations in two dimensions (2D) are usually
strong, positioning the system in the strongly correlated regime, for
which a proper treatment of interaction effects is of paramount
importance, in order to understand various experiments and physical
phenomena at a \emph{quantitative} level.

In the absence of long range order \`a la the Mermin-Wagner theorem
\cite{Mermin1966PRL}, phase transition in 2D is
in general of the Berezinskii \cite{berezinskii1972}, Kosterlitz and
Thouless (BKT) \cite{kosterlitz1973} type. Indeed, 
superfluid transitions in ultracold atomic Bose gases have been
experimentally found to exhibit a strong BKT nature in 2D
\cite{Hadzibabic2006N,Clade2009PRL,Tung2010PRL}.
BKT transitions have been of wide interest since the most important
high $T_c$ superconductors are quasi-2D layered materials
\cite{Kosterlitz2}, for which the low dimensionality is favorable for
the formation of the wide-spread pseudogap phenomena, which are
arguably attributable to strong fluctuations
\cite{Loktev2001PR,chen2024RMP}. It is also the model commonly used
for describing the superconducting transition in thin films
\cite{Hetel2007,xue2013SSC}. Recent exciting discoveries of (quasi-)2D
superconductors
\cite{Bovzovic2016N,Agterberg2017PRL,Hsu2017NC,Cao2018N} have added to
the interest in 2D phase transitions.  The BKT nature of the
superfluid transition in 2D atomic Fermi gases has also been
experimentally confirmed \cite{Ries2015PRL,Murthy2015PRL}. There are
also, albeit not many, theoretical studies of the BKT superfluid
transition in a 2D Fermi system for both weak and strong pairing
interactions, e.g., Ref.~\cite{Wang2020NJP,Shi2024}.

In this paper, we report our study on the important yet often
neglected effect of the particle-hole fluctuations on 2D Fermi gas
superfluidity throughout the BCS-BEC crossover, to which similar
studies has not been reported in the literature
\cite{Shi2024}. Similar to its 3D counterpart, we find that
particle-hole fluctuations lead to a screening to the pairing
interaction, causing a shift of the superfluid transition curve toward
the BEC regime as a function of the pairing strength. This
substantially suppresses $T_c$ and the gap in the BCS and unitary
regimes, and has an important physical significance when comparing
between theory and experiment.
 
It is important to note that there has been a historical controversy
surrounding 2D fermionic superfluids, dating back to Kosterlitz and
Thouless \cite{kosterlitz1973}, which primarily concerns observable
signatures and the applicability of BKT physics. It is not
straightforward to apply the BKT theory based on the XY model to
fermionic superfluids. In the XY model, in which the superfluid
density $n_s/m$ provides the phase stiffness, it is the phase
fluctuations of the vortex type that dominate the destruction of the
quasi-long range superfluid order and drive the system into a
disordered normal state. Unlike the superfluids of true bosons,
however, the superfluid density is also suppressed by the
pair-breaking Bogoliubov quasiparticle excitations as the temperature
$T$ increases. In addition, both the density and the mass of the
bosons (i.e., fermion pair) now depend on the temperature and the
pairing interaction strength, except in the deep BEC regime.
Therefore, computing the superfluid density using a mean-field
approximation would yield a constant value $n_s/m=n/m$ at $T=0$ (or $n/4m$ from the boson point of view, with $n_\text{B} = n/2$ and $M_\text{B} = 2m$),
independent of the interaction strength. (Following standard notations, here $n$, $m$, $n_\text{B}$,  $M_\text{B}$ and $n_s$ denote total fermion number density, fermion mass, Cooper pair number density, Cooper pair mass, and superfluid number density, respectively.) At finite $T$, the pair-breaking effect due to quasiparticle excitations can be partly accounted for via solving the mean-field BCS gap equation, leading to a mean-field result of $(n_s/m)^\text{BCS}$. However, not being able to properly take care of the effective pair mass and pair number density  is thus expected to give
rise to an overestimate of the superfluid transition temperature
$T_\text{BKT}$. Indeed, the fact $T_\text{BKT}$, as measured experimentally in
2D atomic Fermi gases, varies with interaction strength, reflects that
both $n_\text{B}$ and $M_\text{B}$ vary with the interaction.

The pair number density at given temperature and interaction is normally governed by the pair dispersion. The latter, or equivalently pair mass, can, in principle, be extracted from the pair propagator.
However, there are not many calculations of the fermionic $T_\text{BKT}$ in the literature \cite{botelho2006vortex,Bauer2014PRL,Bighin2016PRB,Mulkerin2017PRA}, partly due to the inapplicability of the simple XY model directly to the Fermi system. The fact that  $(n_s/m)^\text{BCS}$ decreases with temperature below $T_\text{BKT}$ in the BCS regime manifests that amplitude fluctuations play an important role. Therefore, the superfluid transition in a 2D Fermi gas  has a much richer physics than the simple XY model can capture.
To account  for the finite temperature and interaction effects, Wu et al \cite{Wu2015PRL} and Wang et al
Ref.~\cite{Wang2020NJP} determined $T_\text{BKT}$ using the critical
phase space density criterion,  based on a quantum Monte Carlo simulation \cite{Prokofev2001PRL} along with experimental support \cite{Murthy2015PRL}. A pairing fluctuation theory
\cite{chen1998PRL} that was developed to address the pseudogap
phenomena in 3D superconductors was used to determine the effective pair number
density $n_\text{B}$ and pair mass $M_\text{B}$, which reflect the
effects of both amplitude and phase fluctuations, governed by temperature and interaction strength. Nevertheless, these earlier works \cite{Wang2020NJP,chen1998PRL} considered only the particle-particle channel of the $T$-matrix.

It has been known that particle-hole fluctuations may play an
important role in the 3D superfluid behavior, despite that they are often
neglected in theoretical treatments of superconductivity
\cite{SchriefferBook}.  Gor'kov and Melik-Barkhudarov (GMB) \cite{GMB} first
found that the lowest order particle-hole fluctuations may reduce both
$T_c$ and the zero temperature gap $\Delta_0$ by a factor of 0.44 in a
BCS superconductor.
The study of the GMB effect at the lowest order were extended
to atomic Fermi gases in the continuum \cite {Heiselberg2000PRL} or an
optical lattice \cite{Kim2009PRL}. The effect beyond the lowest order,
by including the full particle-hole $T$-matrix in a ladder
approximation has been studied without \cite{Yu2009PRA} and with
\cite{Chen2016SR} the self-energy feedback.

In 2D, there have been no similar studies in the context of BCS-BEC
crossover in the literature, however, except for the lowest order
particle-hole contributions in the BCS limit~\cite{Petrov2003PRA}. It
is the purpose of the present work to investigate the effect of
particle-hole fluctuations on the  Fermi gas superfluidity in
2D. The low dimensionality further enhances the strong pairing
fluctuations, which are already present when the interaction is
strong. The very BKT nature of the transition requires that there is
already a sizable pairing amplitude, i.e., (pseudo)gap, at
$T_\text{BKT}$. Therefore, this necessitates the self-consistent
inclusion of the self-energy feedback in (both the particle-particle
and) the particle-hole $T$-matrices.
Here we incorporate the particle-hole channel contributions in the
pairing fluctuation theory \cite{chen1998PRL,Chen1999PRB,Chen2016SR},
and thus go beyond previous studies \cite{Wu2015PRL,Wang2020NJP}.
This pairing fluctuation theory includes self-consistently the
finite-momentum pairing fluctuations in the single-fermion self
energy, and has successfully addressed multiple high $T_c$
\cite{chen1998PRL,Chen1999PRB,Chen2001PRB,ReviewLTP-Full} and atomic
Fermi gas experiments
\cite{chen2005PR,Kinast2005S,Chen2009PRL,FrontPhys}. In particular, it
features naturally a pseudogap in 3D unitary Fermi gases, which has
been unequivocally corroborated by a recent experiment
\cite{li2024nature}.
Following the previous work in 3D \cite{Chen2016SR}, we show that the
particle-hole $T$-matrix serves as a renormalized pairing interaction,
which is to appear in the $T$-matrix in the usual particle-particle
channel.  We find that the particle-hole fluctuations lead to a
screening to the pairing interaction, and the inverse pairing
interaction is effectively shifted by the temperature-dependent,
angular-averaged (at two different levels) particle-hole
susceptibility $\langle\chi_\text{ph}\rangle$, which approaches a
negative constant, $-m/2\pi$, in the BCS limit and increases gradually
in the crossover regime toward zero in the BEC limit.  In result, the
particle-hole channel shifts the $T_\text{BKT}$ curve towards the BEC
regime as a function of pairing strength.  Furthermore, comparison
shows that the inclusion of particle-hole channel leads to a better
overall agreement between our calculated $T_\text{BKT}$ with the results from
experiment and quantum Monte Carlo simulations.

\section{Theoretical Formalism}

\subsection{Overview of the Pair Fluctuation Theory without the Particle-Hole Channel}

To be self-contained, we first recapitulate the pairing fluctuation
theory \cite{chen1998PRL,Chen1999PRB,Wu2015PRL} without including the
particle-hole channel. This serves as a basis, on top of which the
particle-hole channel effect is built. Note that here we will consider
only the formalism without the superfluid order parameter,
tailored for 2D in the absence of a true long
range order.

We consider a 2D Fermi gas with a short-range $s$-wave attractive
interaction $V_{\mathbf{k},\mathbf{k}^{\prime}} = U <0$, as described
by a generic grand canonical Hamiltonian \cite{Chen1999PRB}, which
includes pairing with a finite center-of-mass momentum $\mathbf{q}$.  The free
fermion dispersion is given by $\xi_{\mathbf{k}} =
\epsilon_{\mathbf{k}} - \mu \equiv \mathbf{k}^2/2m - \mu$, with
chemical potential $\mu$. The Fermi momentum is $\hbar k_\text{F} =
\sqrt{2\pi n}\hbar $, and Fermi energy $E_\text{F} \equiv
k_\text{B}T_\text{F} = \hbar^2 k_\text{F}^2/2m$. As usual, we shall
use the natural units, and set $\hbar = k_\text{B}=1$ and the volume
to unity \cite{chen1998PRL}.

The fermions acquire self energy via pair binding and unbinding. Our
approximated equations, derived via an equations of motion approach \cite{Kadanoff1961PR,ChenPhD}, is often
cast into a $T$ matrix formalism, with a mix of bare Green's function
$G_0(K)=(\mathrm{i}\omega_{l}-\xi_{\mathbf{k}})^{-1}$ and full Green's function $G(K)$ in the pair susceptibility
$\chi(Q) = \sum_K G(K)G_0(Q-K)$. Here we use the four-momentum
notation, $K \equiv(\mathrm{i} \omega_n, \mathbf{k})$ and $Q
\equiv(\mathrm{i} \Omega_l, \mathbf{q})$, $\sum_K \equiv T \sum_{n,
  \mathbf{k}}$ and $\sum_Q \equiv T \sum_{l, \mathbf{q}}$, with
$\omega_n$ and $\Omega_l$ being Matsubara frequencies for fermions and
bosons, respectively \cite{fetter}. 
The $T$-matrix $t(Q)$ is given by $t(Q)=U/[1+U \chi(Q)]$, with the self
energy \[ \Sigma(K)=\sum_{Q}t(Q)G_0(Q-K). \] 

The Thouless criterion (for superfluid transition in 3D) requires 
$t^{-1}(Q = 0)=U^{-1}+\chi(0)=0$. In order to accommodate the absence
of long range order in 2D, we generalize the Thouless criterion to
allow for a very small but finite pair chemical potential $
\mu_\text{p} $, $ U^{-1}+\chi(0) = t^{-1}(0)  \propto\mu_\text{p}$.
In this way, $t(Q)$ is highly peaked
around $Q=0$.  Thus the main contribution to $\Sigma(K)$ comes from
the vicinity of $Q=0$, leading to a pseudogap approximation,
 $\Sigma(K) \approx -\Delta^2 G_0(-K)$, where we define the
pseudogap as $\Delta^2 =-\sum_{Q}t(Q)$.
With this approximation, the full Green's function takes a simple BCS-like form and is given by
\begin{equation*}
    G(K)=\frac{u_{\mathbf{k}}^{2}}{\mathrm{i}\omega_{n}-E_{\mathbf{k}}}+\frac{v_{\mathbf{k}}^{2}}{\mathrm{i}\omega_{n}+E_{\mathbf{k}}}\,,
\end{equation*}
where $E_{\mathbf{k}}=\sqrt{\xi_{\mathbf{k}}^2+\Delta^2}$, 
$u_\mathbf{k}^2 = (1+\xi_\mathbf{k}/E_{\mathbf{k}})/2$, 
and $v_\mathbf{k}^2 = (1-\xi_\mathbf{k}/E_{\mathbf{k}})/2$.
%
This leads to a 
BCS-like gap equation,
\begin{equation}
  \label{eq:gap}
  a_0 \mu_\text{p} = \sum_{\mathbf{k}}\left[\frac{1-2 f(E_{\mathbf{k}})}{2 E_{\mathbf{k}}} - \frac{1}{2\epsilon_\mathbf{k} + \epsilon_\text{B}}\right]\,,
\end{equation}
where 
$a_0$ is the coefficient of the linear $\Omega$ term in the Taylor expansion of the inverse $T$-matrix. 
Here the gap equation has been regularized via $U^{-1} = - \sum_{\mathbf{k}}  1/(2 \epsilon_{\mathbf{k}}+\epsilon_{\mathrm{B}})$, 
where the two-body binding energy $\epsilon^{}_\text{B} = 1/ma^2_\text{2D}$ with the 2D scattering length $a^{}_\text{2D}$ \cite{Levinsen2015}. 
The fermion number constraint $n = 2\sum_K G(K)$ yields 
\begin{equation}
  \label{eq:eqn}
  n = \sum_{\mathbf{k}}\left[1-\frac{\xi_{\mathbf{k}}}{E_{\mathbf{k}}}+2 \frac{\xi_{\mathbf{k}}}{E_{\mathbf{k}}} f(E_{\mathbf{k}})\right]\,,
\end{equation}
where $f(x)$ is the Fermi distribution function. 

To extract the pair dispersion, one  Taylor-expands $t^{-1}$ near $Q=0$ and analytically continues to  real frequency, with
 $(\mathrm{i}\Omega_l \rightarrow \Omega+\mathrm{i}0^+)$, so that
\begin{equation} t^{-1}(\Omega,\mathbf{q}) \approx a_1\Omega^2+a_0(\Omega-\Omega_\mathbf{q}^0+\mu_\text{p}),\end{equation}
where $\Omega_{\mathbf{q}}^0 = \mathbf{q}^2 / 2M_\text{B}$.
Consequently, the definition of the pseudogap
yields
\begin{equation}
   \label{eq:pg}
 n_\text{B}\equiv  a_0 \Delta^2 = \sum_{\mathbf{q}}\left[1+4\frac{a_1}{a_0}(\Omega_{\mathbf{q}}^0-\mu_\text{p})\right]^{-1/2} b(\Omega_{\mathbf{q}}),
\end{equation}
where $b(x)$ is the Bose distribution function 
and $\Omega_{\mathbf{q}}=\left[\sqrt{a_0^2+4a_0a_1(\Omega_{\mathbf{q}}^0-\mu_\text{p})}-a_0\right]/2a_1$ represents the pair dispersion. 
The coefficients $a_0$, $a_1$, $M_\text{B}$ are determined via the expansion process. 
Except in the weak coupling BCS regime, the $a_1$ term serves as a small quantitative correction and can often be neglected, so that $\Omega_{\mathbf{q}}\approx \Omega_{\mathbf{q}}^0-\mu_\text{p}$. The way Eq.~(\ref{eq:pg}) yields $n_\text{B}$ can be seen from the relation $n_\text{B} = n/2 -\sum_\mathbf{k} f(\xi_\mathbf{k}^{})$ \cite{ChenPhD}.

\subsection{BKT Criterion}

The Nelson-Kosterlitz condition (NKC) provides a criterion for determining $T_\text{BKT}$. However, there are issues with this condition. First of all, due to the Galilean translational invariance,  one has the zero $T$ superfluid density $n_s(T=0) = n$, independent of the pairing strength. Furthermore, from the fermionic perspective, the fermion mass $m$ is a constant. Hence it is hard to imagine that $n_s(T_\text{BKT})/m$ would be able to properly capture the interaction effect. While it is often assumed $M_\text{B}=2m$ in a NKC-based calculation, 
however, the correct $M_B$ should  necessarily be renormalized by the interactions. Therefore, it is reasonable to suspect that NKC is not quantitatively reliable for a fermionic system.

Here we choose the criterion of critical phase space density ${\cal D}_B$ to determine  $T_\text{BKT}$, as established by Prokof'ev et al via quantum Monte Carlo (QMC) simulations of a Bose gas \cite{Prokofev2001PRL}. When approached from  high temperature
\cite{kosterlitz2013,Hadzibabic2006N}, the bosonic BKT transition
occurs when $\mathcal{D}_{\text{B}}(T)$ reaches a critical value
$\mathcal{D}_{\text{B}}(T_\text{BKT})$ such that the boson wavefunctions start to overlap with each other \cite{Prokofev2002PRA}. Hence we have
\begin{equation}
  \label{eq:BKT}
  \frac{n^{}_{\text{B}}}{M_{\text{B}}}=\frac{{\cal D}_\text{B}}{2 \pi} T_{\text{BKT}}\,,
\end{equation}
where $n_\text{B}$ and $M_{\text{B}}$ are determined as described
above. In the deep BEC regime, we have $n_\text{B}=n/2$ and $M_\text{B} = 2m$, and
hence the asymptotic behavior $T_{\text{BKT}}=T_\text{F}/2{\cal D}_\text{B}$. This would reduce to the NKC when   ${\cal D}_\text{B}=4$.

The critical phase space density ${\cal D}_\text{B}$ has a very weak loglog dependence on $a_\text{2D}$, as given by $ {\cal D}_\text{B} = \ln (C/\tilde{g})$, with $C\approx 380$ \cite{Prokofev2001PRL}. Instead of using the expression for $\tilde{g}$ in terms of potential range \cite{Prokofev2001PRL} or pair size \cite{Petrov2003PRA}, we take it directly from the experiment  by fitting the  data with the right form of function, which yields a perfect fit,
\begin{equation}
\tilde{g} = \frac{5.03085}{1.23629 - \ln (k_\text{F}a_\text{2D})}.
\label{eq:g}
\end{equation}
See App.~\ref{App:D_B} for details.  The corresponding $ {\cal D}_\text{B}$ is given by the blue curve in Fig.~\ref{fig:g}. The fit also reveals that there exists a quantitative difference between $a_\text{2D}$ and either the potential radius or the pair size.

The fit Eq.~(\ref{eq:g}) is not to be applicable in the unitary and BCS regime. Since the relation $ {\cal D}_\text{B} = \ln (C/\tilde{g})$ no longer works well for the next available data point $\tilde{g}=7.55$ toward the unitary and BCS regimes \cite{Murthy2015PRL}, we set 4.9 as the lower bound for ${\cal D}_\text{B}$ in the entire BCS-BEC crossover \footnote{Note that the concrete value of  ${\cal D}_\text{B}$ will have a slight quantitative impact on $ T_{\text{BKT}}$ in the unitary regime. Such an impact becomes negligible in the BCS regime where both $(n_s/m)$ and $n_\text{B}/M_{\text{B}}$ decrease rapidly with $T$ around $T_{\text{BKT}}$ so that $T_{\text{BKT}}\approx T_c^\text{MF}$. }.

At $T=0$ where $\mu_\text{p}=0$, Eqs.~(\ref{eq:gap}) and
(\ref{eq:eqn}) give $\mu = E_\text{F}-\epsilon^{}_\text{B}/2$ and
$\Delta = \sqrt{2E_\text{F}\epsilon_\text{B}}$ \cite{Randeria1990PRB}.
At finite temperature, the summation in Eq.~(\ref{eq:pg}) analytically
leads to
$\frac{a_0}{a_1}-\sqrt{\frac{a_0}{a_1}(\frac{a_0}{a_1}-4\mu_\text{p})}
= 2T \ln(1-e^{-\mathcal{D}_{\text{B}}})$, which reduces to
$\mu_\text{p} = T \ln(1-e^{-\mathcal{D}_{\text{B}}})$ in the BEC
regime.  Here use has been made of the relation
$\mathcal{D}_{\text{B}} = 2 \pi n_{\text{B}} / M_\text{B} T $.
Thus $\mu_\text{p}$ is primarily determined by
$\mathcal{D}_{\text{B}}$ at given temperature, indicating that when
the BKT transition occurs $\mathcal{D}_{\text{B}}$ is large enough so
that $\mu_\text{p}$ becomes sufficiently close to zero
\cite{Prokofev2002PRA,Murthy2015PRL,Wu2015PRL,Wang2020NJP}.  Moreover,
in the BEC regime, a large $\mathcal{D}_{\text{B}}$ provides a sharp
peak distribution of the bosonic number density at zero momentum via
$b(-\mu_\text{p})=e^{\mathcal{D}_{\text{B}}}-1$, signaling a
quasi-condensation \cite{Wu2015PRL}. Alternatively, when ${\cal
  D}_\text{B}(T)$ becomes large enough, the wave functions of neighboring
pairs start to overlap with each other, and hence help to establish
quasi-long-range phase coherence.

\subsection{Contributions of the Particle-Hole Channel}

Following Ref.~\cite{Chen2016SR}, we introduce the contribution of the
particle-hole channel, by renormalizing the pairing strength in
$t(Q)$, which leads to a new full $T$-matrix $t_\text{2}(Q)$ that
includes both particle-particle and particle-hole contributions.  The
expression for $t_\text{2}(Q)$ is formally given by
\begin{equation*}
  t_\text{2}(Q) = \frac{1}{t^{-1}_\text{ph}(K + K' - Q) + \chi(Q)}\,,
\end{equation*}
which self-consistently includes the self-energy feedback, with $K,K'$
being the external fermion momentum.  Here the particle-hole channel
$T$ matrix $t^{-1}_\text{ph}(Q') = U^{-1} + \chi_{\text{ph}}(Q')$
describes the particle-hole scattering, and the particle-hole
susceptibility $\chi_{\text{ph}}(Q') = \sum_{K} G(K) G_0(K-Q')$, where
the particle-hole momentum $Q'=K + K' - Q$.  Moreover, assuming that
the fermions near the Fermi surface dominate the particle-hole channel
contributions, we replace the particle-hole susceptibility with an
average $\langle{\chi_\text{ph}\rangle}$ on or near the Fermi surface,
where the frequency part of the particle-hole susceptibility is set to
0 with $\mathrm{i}\Omega'_n = 0$ \cite{gor1961}, leading to a zero
imaginary part of $\chi_{\text{ph}}(Q')$ and therefore a purely real
$\langle\chi_{\text{ph}}\rangle$ \cite{Chen2016SR}.

We have proposed two methods for averaging $\chi_{\text{ph}}(Q')$,
referred to as level 1 and level 2, respectively.  The level 1 average
involves an on-shell and elastic scattering on the Fermi surface, with
$|\mathbf{k}|=|\mathbf{k}'|=k_\mu = \sqrt{2m\max(\mu,0)}$, where the
momentum part of $\chi_{\text{ph}}(Q')$ is determined by
$|{\mathbf{q}'}|=\left|\mathbf{k}+\mathbf{k}^{\prime}\right|=
k_\mu\sqrt{2(1+\cos \theta)}$.  Here $\chi_{\text{ph}}(Q')$ is
averaged over scattering angles $\theta$ (between $\mathbf{k}$ and
$\mathbf{k}^{\prime}$).  This averaging process, focusing solely on
the Fermi surface, is commonly used in the literature on the studies
of induced interactions \cite{Yu2009PRA}.  In contrast, the level 2 average considers
that the states within the energy range $\xi_\mathbf{k} \in
[-\min(\Delta, \mu),\Delta]$ of a typical s-wave superconductor are
most significantly affected by pairing (for $\mu > 0$).  Hence, for
the level 2 average, while keeping the on-shell and elastic
scattering assumption, the average is performed over a range of $|\mathbf{k}|$
such that the quasi-particle energy $E_{\mathbf{k}} \in\left[\min
  (E_{\mathbf{k}}), \min (\sqrt{E^2_{\mathbf{k}}+\Delta^2})\right]$,
where $\min (E_{\mathbf{k}})=\Delta$ if $\mu>0$, or $\min
(E_{\mathbf{k}})=\sqrt{\mu^2+\Delta^2}$ if $\mu<0$.

Then with this frequency and momentum independent $\langle{\chi_\text{ph}\rangle}$, 
the new full $T$-matrix $t_\text{eff}(Q)$ reads
\begin{equation}
  \label{teff}
  t_\text{eff}(Q) = \frac{1}{U^{-1} + \langle\chi_{\text{ph}}\rangle + \chi(Q)}\,.
\end{equation}
The gap equation with the particle-hole channel effect is modified into 
\begin{equation}
  \label{eq:gapph}
  a^{}_0 \mu_\text{p} =  \langle\chi_{\text{ph}}\rangle + \sum_{\mathbf{k}}\left[\frac{1-2 f(E_{\mathbf{k}})}{2 E_{\mathbf{k}}} - \frac{1}{2\epsilon^{}_\mathbf{k} + \epsilon^{}_\text{B}}\right]\,,
\end{equation}
while the other equations remain unchanged.

Equations (\ref{eq:eqn}), (\ref{eq:pg}), and (\ref{eq:gapph}) form a closed set of self-consistent equations, 
which can be used to solve for $(\mu, \Delta, \mu_\text{p})$, along with $T_\text{BKT}$ via the BKT criterion given by Eq.~(\ref{eq:BKT}). 

Throughout the BCS-BEC crossover,  $\langle\chi_{\text{ph}}\rangle$ remains negative as a function of the coupling strength.
From Eq.~(\ref{eq:gapph}), the particle-hole channel constitutes a renormalization of the pairing interaction, with a net effect given by replacing $1/U$ with $ 1/U_\text{eff}\equiv 1/U + \langle\chi_{\text{ph}}\rangle$. Diagrammatically, this amounts to replacing the bare interaction $U$ with the full particle-hole $T$-matrix $t_\text{ph}$ in the particle-particle scattering diagrams \cite{Chen2016SR}. Since $\langle\chi_{\text{ph}}\rangle < 0$, one has $|U_\text{eff}|< |U|$. Thus  $U_\text{eff}$ represents a weaker, screened pairing interaction.

\section{Numerical Results and Discussions}

\subsection{Behaviors of the particle-hole susceptibility}

\begin{figure}
\centerline{\includegraphics[clip,width=3.4in]{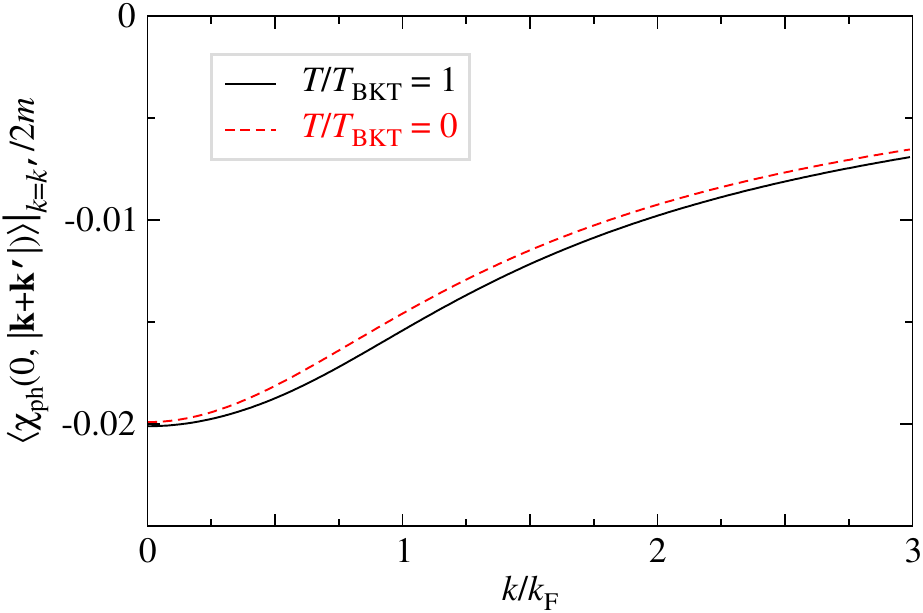}} 
\caption{
  Angular average of  the on-shell particle-hole susceptibility $\langle\chi_{\text{ph}}(0,|\mathbf{k}+\mathbf{k}'|)\rangle/2m$ with $k=k'$ 
  as a function of momentum $k/k_\text{F}$ at unitarity $\ln(k_\text{F} a_\text{2D}) = 0$ for $T=0$ and $T=T_\text{BKT}$, 
  where $T_\text{BKT}/T_\text{F} = 0.079 $ and the corresponding $\Delta$, $\mu$ and $\mu_\text{p}$ are calculated without the particle-hole channel effect.}
\label{fig:chiph}
\end{figure}

First, we present in Fig.~\ref{fig:chiph} the (level 1) angular
average of the particle-hole susceptibility at zero frequency as a
function of momentum $k$, under the on-shell condition
$|\mathbf{k}|=|\mathbf{k}'|=k$.  Here we focus on the unitary case at
$T=T_\text{BKT}$ (black solid line) and zero-temperature (red dashed line), and
$\langle\chi_{\text{ph}}(0,|\mathbf{k}+\mathbf{k}'|)\rangle$ is
calculated using the corresponding solution of $(\Delta,\mu, 
\mu_\text{p})$ at $T_\text{BKT}/T_\text{F} = 0.079$ and
$\ln(k_\text{F} a_\text{2D}) = 0$,  solved in the
absence of particle-hole fluctuations. The slight difference between these two curves reveals 
a weak temperature dependence. Importantly, both curves show a significant momentum dependence, with the amplitude decreasing
monotonically as the momentum rises.
Note that the momentum dependencies in 2D appear distinct from those
in 3D \cite{Chen2016SR}, where
$\langle\chi_{\text{ph}}(0,|\mathbf{k}+\mathbf{k}'|)\rangle$ exhibits
a nonmonotonic momentum dependence at low $T$ for the unitary case.

\begin{figure}
\centerline{\includegraphics[clip,width=3.4in]{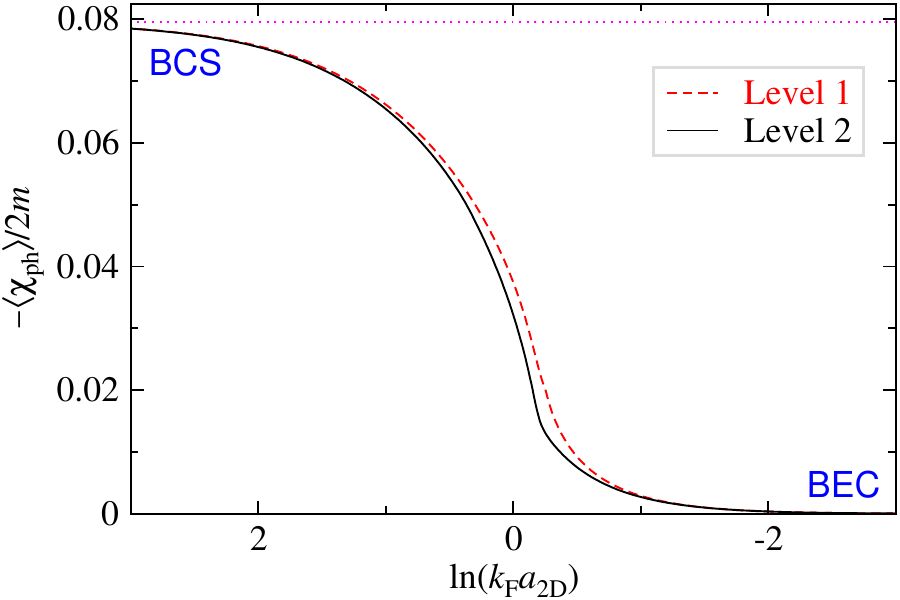}} 
\caption{ $-\langle\chi_{\text{ph}}\rangle / 2m$ at $T=T_\text{BKT}$,
  averaged at both level 1 (red dashed) and level 2 (black solid line),
  as a function of $\ln(k_\text{F}a_\text{2D})$ throughout BCS-BEC
  crossover.  The magenta dotted line indicates the BCS limit value,
  $1/4\pi$, given by $-\langle\chi_{\text{ph}}\rangle = {m}/{2\pi}$.}
\label{fig:chiphTc}
\end{figure}

Shown in Fig.~\ref{fig:chiphTc} is $-\langle\chi_{\text{ph}}\rangle$
at $T=T_\text{BKT}$ as a function of $\ln(k_\text{F}a_\text{2D})$
throughout the BCS-BEC crossover, averaged at both level 1 (red
dashed) and level 2 (black solid curve), along with its BCS limit
$1/4\pi\approx 0.0796$, given by
$\langle\chi_{\text{ph}}\rangle=-{m}/{2\pi}$.  The value of
$-\langle\chi_{\text{ph}}\rangle / 2m$ decreases monotonically with
increasing interaction strength and exhibits an exponential behavior
in the deep BEC regime for $\ln(k_\text{F} a_\text{2D}) < -2$. As the
gap diminishes, the average at both levels converges in the BCS limit.
However, in the crossover regime, the two levels differ significantly,
with a smaller absolute value for the level 2 average. This can be
understood from Fig.~\ref{fig:chiph}; compared with level 1 averaging,
level 2 averaging involves a range of $k$'s so that the larger $k$
contributions dominate due to its larger phase space area, leading to
a smaller absolute value of the average. In other words, level 1
averaging on the Fermi surface only leads to a significant
over-estimate of the particle-hole contributions in the unitary
regime.

The asymptotic behavior of $\langle\chi_{\text{ph}}\rangle$ can be
readily solved analytically in both the BCS and the BEC limits.  In
the weak coupling limit, where $\ln(k_\text{F} a_\text{2D})
\rightarrow \infty$, $\Delta \rightarrow 0$, and $ T_\text{BKT}
\rightarrow 0$, the two levels of average converge to
$\chi_{\text{ph}}(Q') \approx \sum_{K} G_0(K) G_0(K-Q')$. Under the
on-shell condition $\xi_{\mathbf{k}}=\xi_{\mathbf{k}-\mathbf{q}'}$,
the integrand becomes the derivative of the Fermi function and one
readily obtains
\begin{equation}
  \chi_{\text{ph}}(0,\mathbf{q}') = \sum_\mathbf{k}\frac{f(\xi_{\mathbf{k}})-f(\xi_{\mathbf{k}-\mathbf{q}'})}{\xi_{\mathbf{k}}-\xi_{\mathbf{k}-\mathbf{q}'}} \approx -\frac{m}{2\pi} = \langle\chi_{\text{ph}} \rangle\,.
  \label{eq:chiph}
\end{equation} 
We emphasize that this result is independent of the density, due to
the constant density of states in 2D. This is to be contrasted with
its 3D counterpart, which is proportional to the Fermi momentum
$k_\text{F}$. In the strong coupling limit, where $\ln(k_\text{F}
a_\text{2D}) \rightarrow -\infty$, we have $|\mu| \gg \Delta \gg
E_\text{F}$, so that $E_\mathbf{k} \approx \xi_\mathbf{k} \approx
|\mu|$. Thus particle-hole fluctuations are exponentially suppressed,
so that $\langle\chi_{\text{ph}}\rangle$ approaches zero.

\subsection{Effect of the particle-hole channel on  BKT transition}

\begin{figure}
\centerline{\includegraphics[clip,width=3.4in]{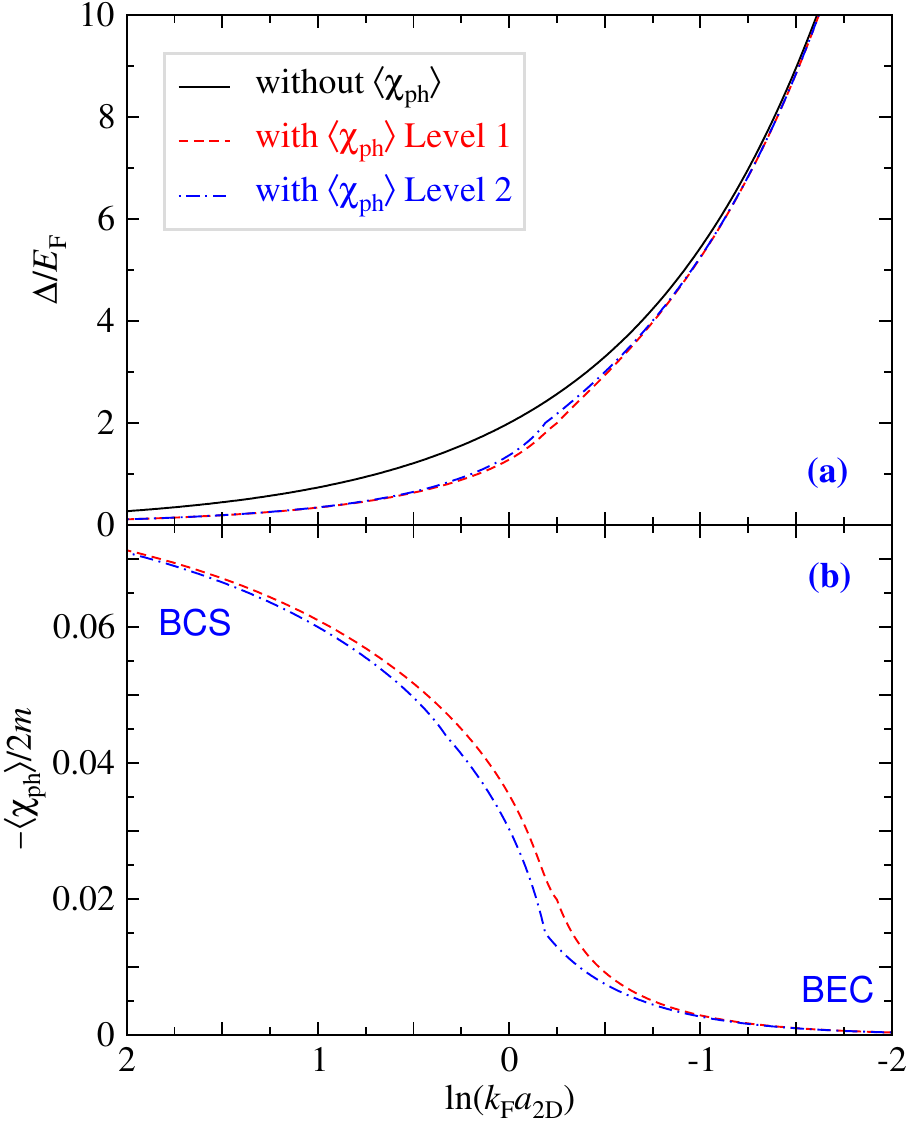}} 
\caption{ (a) Effect of the particle-hole channel contributions on
  $\Delta$, along with (b) the corresponding
  $-\langle\chi_{\text{ph}}\rangle / 2m$, at $T=0$, as the function of
  $\ln(k_\text{F}a_\text{2D})$ throughout the BCS-BEC crossover. Shown
  are results without (black solid curve) and with particle-hole
  channel contributions averaged at level 1 (red dashed) and level 2
  (blue dot-dashed curve).  }
\label{fig:chiphT0}
\end{figure}

\begin{figure*}
\centerline{\includegraphics[clip,width=6.8in]{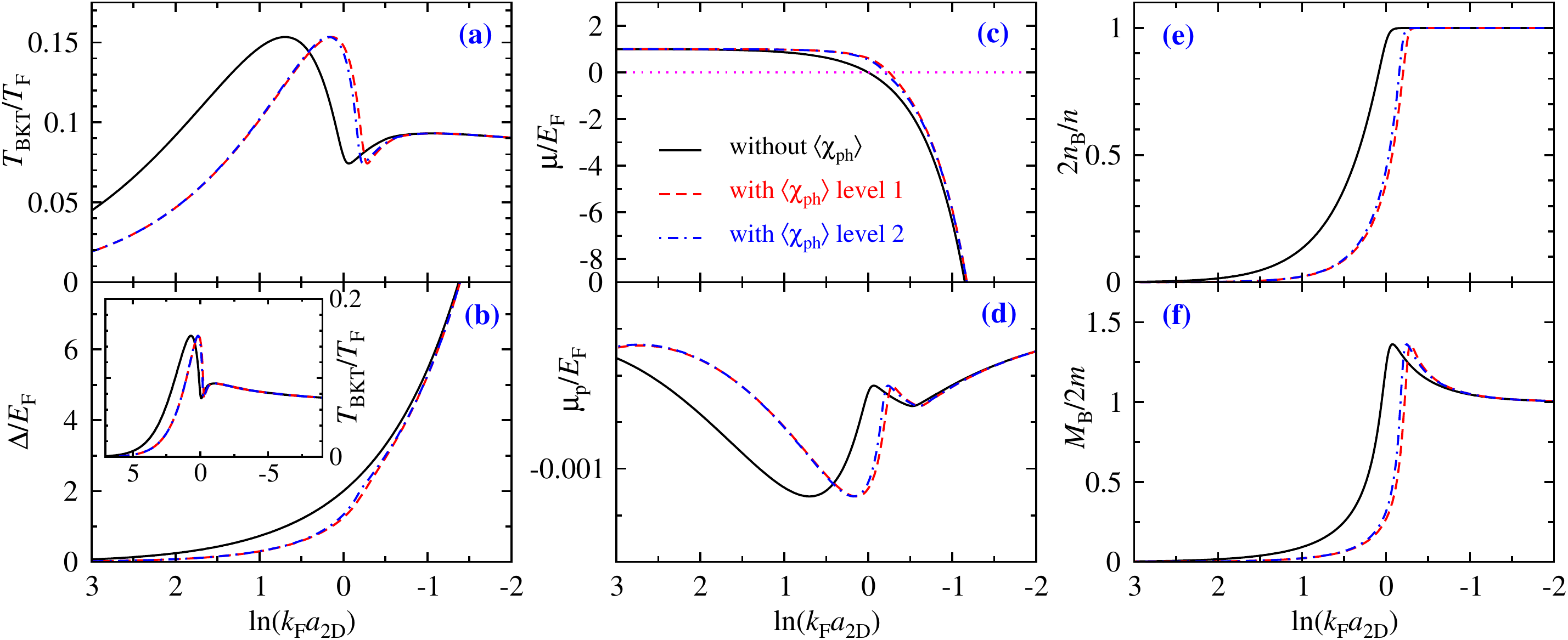}}
\caption{ Effect of the particle-hole channel on (a) $T_\text{BKT}$,
  (b) $\Delta$, (c) $\mu$, (d) $\mu_\text{p}$, (e) $n_\text{B}$ and
  (f) $M_\text{B}$ as a function of $\ln(k_\text{F}a_\text{2D})$, with
  the average $\langle\chi_{\text{ph}}\rangle / 2m$ calculated at
  level 1 (red dashed) and level 2 (blue dot-dashed curves),
  respectively.  They should be compared with the results without the
  particle-hole contributions (black solid curves). Shown in the inset
  of (b) is $T_\text{BKT}$ over a broader range. The difference
  between levels 1 and 2 is not dramatic.}
\label{fig:Tc}
\end{figure*}

In this section, we explore the impact of the particle-hole channel on
the pairing phenomena and the BKT transition of a 2D Fermi gas.  We
begin by examining the behavior of the pairing gap $\Delta$ at zero
temperature with and without the particle-hole channel contributions.
Plotted in Fig.~\ref{fig:chiphT0}(a) is $\Delta$ as a function of
interaction, parameterized by $\ln(k_\text{F}a_\text{2D})$. For
comparison, we plot the results both without (black solid line) and
with the particle-hole channel effect, with the particle-hole
susceptibility $\langle{\chi_\text{ph}\rangle}$ averaged at level 1
(red dashed curve) and level 2 (blue dot-dashed curve), respectively.
The corresponding $\langle{\chi_\text{ph}\rangle}$ is shown in
Fig.~\ref{fig:chiphT0}(b). Compared to Fig.~\ref{fig:chiphTc}, the
difference in$\langle\chi_{\text{ph}}\rangle $ is larger at $T=0$ than
at $T=T_\text{BKT}$. For all cases, $\Delta$ increases monotonically
from BCS to BEC, as expected.  A substantial reduction of $\Delta$ by
the particle-hole fluctuations occurs in the crossover and BCS
regimes.  In the BEC regime, $\langle{\chi_\text{ph}\rangle}$
approaches zero, rendering the particle-hole channel effect
negligible.  Consistent with Fig.~\ref{fig:chiphT0}(b), level 2
averaging showed a weaker particle-hole effect, and thus a smaller
reduction of $\Delta(T=0)$.  Note that in Fig.~\ref{fig:chiphT0}(b),
there exists a kink where $\mu=0$ in $\langle{\chi_\text{ph}\rangle}$
for both levels of averaging, mainly because the Fermi surface shrinks
to zero abruptly with a finite slope as a function of increasing
pairing strength.  The Fermi function in the integrand of
$\langle{\chi_\text{ph}\rangle}$ becomes a step function at zero $T$,
resulting in a Dirac delta function in the derivative of the integrand
of $\langle{\chi_\text{ph}\rangle}$ and hence a slope discontinuity
when crossing $\mu=0$ (See Appendix \ref{sec:AppA} for details). This
slope discontinuity becomes more prominent in the level 2 average,
since the range for $k < k_\mu$ in the average also shrinks to zero
abruptly when $\mu= 0$ \footnote{It should be noted that $\mu=0$
occurs at $\ln (k_Fa_\text{2D}) \approx -0.25 $ and -0.175 for level 1
and 2 averaging, respectively, as they are calculated with the
self-consistent solutions of $T_\text{BKT}$.}.

For the effect of the particle-hole channel on the behavior of the BKT
transition temperature $T_\text{BKT}$, we first analytically estimate
the ratio between the two BKT transition temperatures in the BCS limit
with and without the particle-hole channel at the same coupling
strength, where the latter is denoted as $T_\text{BKT}^\text{BCS}$.
Now that $\Delta \rightarrow 0$, we obtain
\begin{equation*}
    \sum_{\mathbf{k}}\left[\frac{1-2 f(\xi_{\mathbf{k}})}{2 \xi_{\mathbf{k}}}-\frac{1}{2 \epsilon_{\mathbf{k}}+\epsilon_\text{B}}\right] = \frac{m}{4\pi} \ln \left( \frac{2e^{2\gamma} \epsilon_\text{B} E_\text{F}}{\pi^2 T^2}\right),
\end{equation*}
where $\gamma \approx 0.5772157$ is Euler’s constant.
Substituting the above relation for the corresponding term in
Eqs.~(\ref{eq:gap}) and (\ref{eq:gapph}), we obtain
\begin{equation*}
  \frac{T_\text{BKT}}{T_\text{BKT}^\text{BCS}} = e^{ 2 \pi \langle\chi_{\text{ph}}\rangle / m} = e^{-1} \approx 0.37 \,,
\end{equation*}
where we have taken into account that $t^{-1}(0)$ and
$t_\text{2}^{-1}(0)$ are sufficiently small so that
$|t^{-1}(0)|,|t_\text{2}^{-1}(0)| \ll
|\langle\chi_{\text{ph}}\rangle|$ in the weak coupling limit.

Next, we present in Fig.~\ref{fig:Tc} the effect of the particle-hole
channel on the evolution of (a) $T_\text{BKT}$, along with (b)
$\Delta$, (c) $\mu$, (d) $\mu_\text{p}$, (e) $n_\text{B}$, and (f)
$M_\text{B}$ at $T_\text{BKT}$ as a function of
$\ln(k_\text{F}a_\text{2D})$.  Shown in the inset is $T_\text{BKT}$
over a broad range of $\ln(k_\text{F}a_\text{2D})$. For comparison,
the results without particle-hole fluctuations are presented as the
black solid curves. Consider the case without particle-hole effects
first. Starting from the weak coupling BCS limit (black curve), as the
interaction strength increases, $T_\text{BKT}$ increases, and then
reaches a maximum around $\ln (k_\text{F}a_\text{2D})=0.7$ in the
intermediate regime, where $\mu_\text{p}$ reaches a minimum
simultaneously.  The fact there appears a maximum reflects that
$n_\text{B}$ rises faster than $M_\text{B}$ in this regime (see
(e,f)). As the interaction strength increases further past the
maximum, $T_\text{BKT}$ decreases and reaches a minimum near unitarity
$\ln (k_\text{F}a_\text{2D})=0$, where $\mu=0$.  Beyond this point,
the system enters the BEC regime, where all fermions pair up with
$2n_\text{B}/n \approx 1$, as shown in Fig.~\ref{fig:Tc}(e).  The
behavior of $T_\text{BKT}$ is then dominated by the shrinking pair
size, as indicated by a decrease in the pair mass $M_\text{B}$ toward
$2m$ (Fig.~\ref{fig:Tc}(f)), reaching the BEC asymptotic behavior
$T_\text{BKT}/T_\text{F}=1/2{\cal D}_\text{B}$ governed by
Eq.~(\ref{eq:BKT}). This decreasing $M_\text{B}$ leads to an increase
and hence a minimum in $T_\text{BKT}$.  The gap $\Delta$ in
Fig.~\ref{fig:Tc}(b) increases consistently with pairing strength. In
comparison, the particle-hole contributions cause a \emph{non-uniform
shift} of all curves toward the BEC regime on the right. This shift is
the largest in the BCS limit, and vanishes in the BEC regime, as
indicated in Fig.~\ref{fig:chiphTc}. It exhibits a strong dependence
on $\ln (k_\text{F}a_\text{2D})$ in the crossover regime. Now the
maximum of $T_\text{BKT}$ occurs closer to unitarity, and the location
for $\mu=0$ is shifted into the previous BEC regime.  All
$T_\text{BKT}$ curves with and without the particle-hole effect
converge to the same BEC asymptotic curve, since the particle-hole
contributions vanish gradually in the BEC regime.  From
Fig.~\ref{fig:Tc}(d), we see a tiny $|\mu_\text{p}| \le 1.2 \times
10^{-3} E_\text{F}$ throughout the BCS-BEC crossover for all
cases. Especially, in the BEC regime, it is uniquely determined by
${\cal D}_\text{B}$. This ensures that $t(Q)$ remains highly peaked at
$Q=0$ and thus validates our pseudogap approximation for the
self-energy.

\begin{figure*}
\centerline{\includegraphics[clip,width=6.8in]{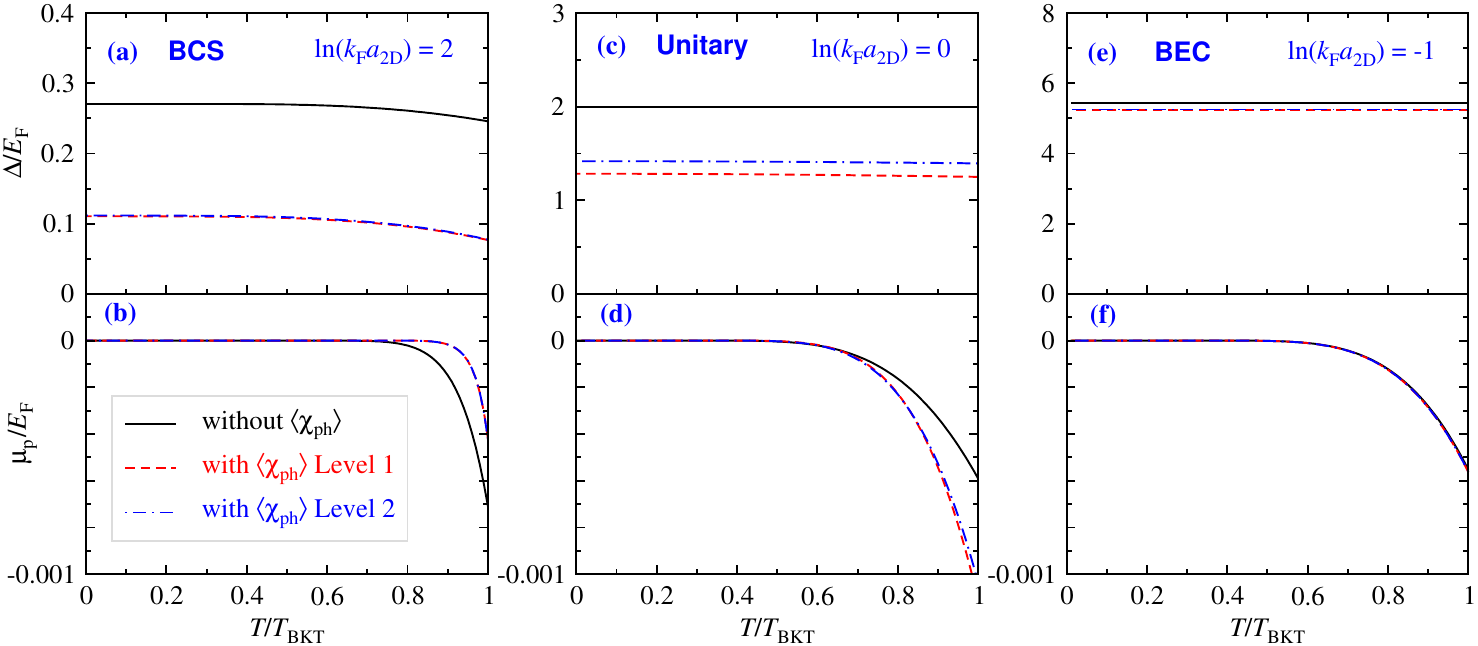}}
\caption{ Effect of the particle-hole channel contributions on
  behaviors of $\Delta$ (top row) and $\mu_\text{p}$ (bottom row), as a
  function of $T/T_\text{BKT}$, with
  $\ln(k_\text{F}a_\text{2D})=2,0,-1$ from left to right for the BCS,
  unitary, and BEC regimes, respectively.  The black solid curves
  represent calculations without the particle-hole channel,
  while the red dashed and green dot-dashed curves include the
  particle-hole channel effect, using $\langle\chi_{\text{ph}}\rangle
  / 2m$ under level 1 and level 2 averaging, respectively.  }
\label{fig:BT}
\end{figure*}

Now we investigate in Fig.~\ref{fig:BT} the evolution of $\Delta$ (top
row) and $\mu_\text{p}$ (bottom row) as a function of reduced
temperature $T/T_\text{BKT}$ in the BCS, unitary, and BEC regimes from
left to right without (black) and with (red and green) the
particle-hole channel effect.  The gap $\Delta$ remains nearly
constant except in the BCS case (a), where a significant decrease can
be discerned near $T_\text{BKT}$.  The pseudogap $\Delta$ with the
particle-hole channel effect is reduced from the counterpart without
the particle-hole effect by a factor of $1/e$ in the BCS limit. This
reduction factor becomes smaller in the unitary and BEC regimes,
consistent with the zero $T$ and $T_\text{BKT}$ gap behaviors shown in
Fig.~\ref{fig:chiphT0}(a) and Fig.~\ref{fig:Tc}(b).  At the same time,
for all cases, $\mu_\text{p}$ increases continuously to zero as $T$
decreases, consistent with $\mu_\text{p}=0$ at $T=0$ for a true
long-range-ordered ground state. Note that below $0.6T_\text{BKT}$,
$\mu_\text{p}$ comes essentially zero. Interestingly, $|\mu_\text{p}|$
decreases much faster with decreasing $T$ in the BCS regime than in
the unitary and BEC regimes. Conversely, should the calculation be
extended to above $T_\text{BKT}$, $|\mu_\text{p}|$ would rise rapidly
with $T$ so that pairs would quickly become irrelevant in this regime.

\subsection{Comparison with different results}

\begin{figure}
\centerline{\includegraphics[width=3.4in]{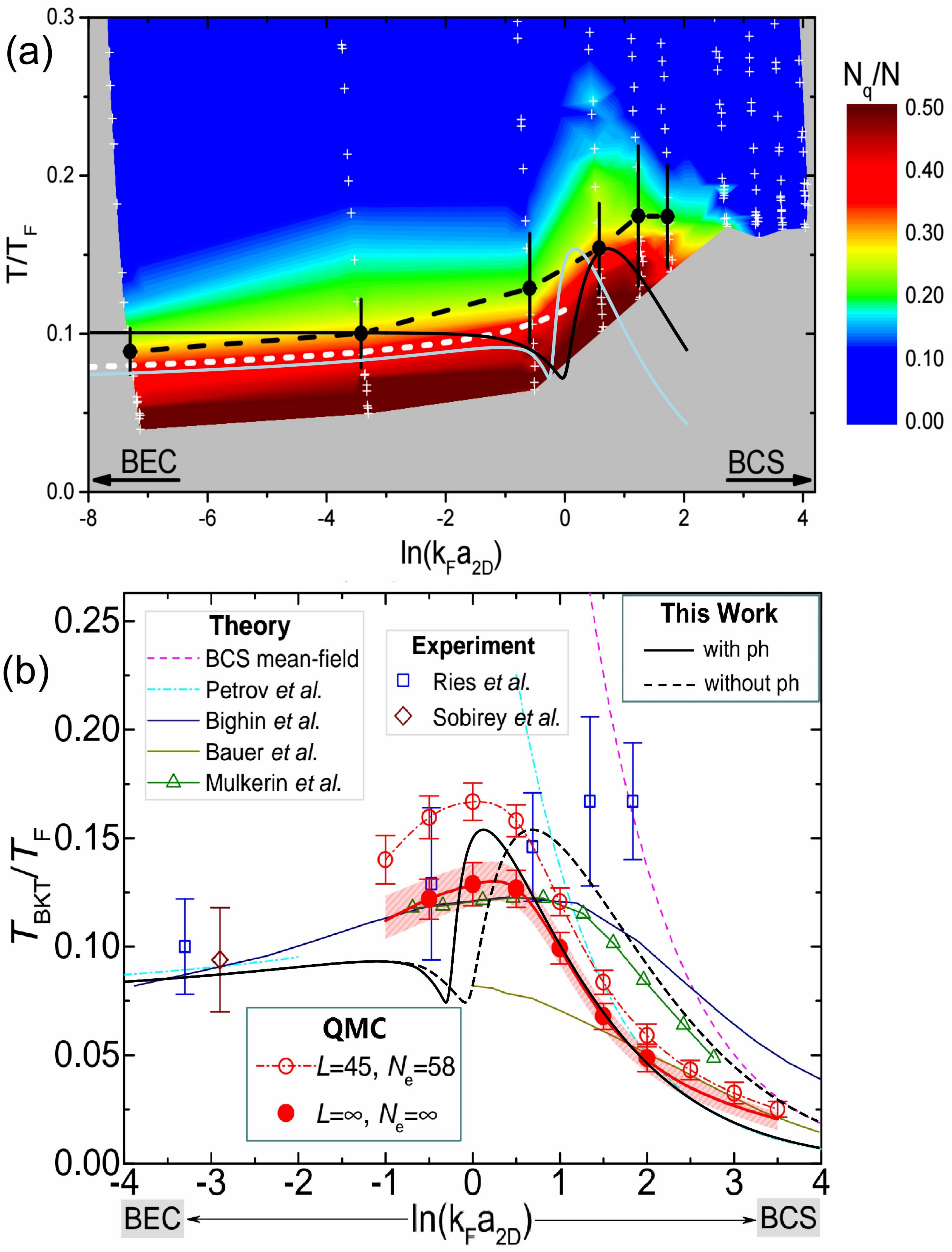}}
  \caption{ Comparison of theoretical calculations for
    $T_\text{BKT}/T_\text{F}$ with experiment
    \cite{Ries2015PRL,Lennart2021} and QMC \cite{He2022PRL} results as
    a function of $\ln(k_\text{F}a_\text{2D})$ throughout the BCS-BEC
    crossover.  (a) Overlay of theoretical $T_\text{BKT}$ without
    (black solid line, with fixed ${\cal D}_\text{B}=4.9$ from
    Ref.~\cite{Wang2020NJP}) and with the particle-hole channel
    calculated with the level 1 average (light cyan solid line), on
    top of the contour plot of experimentally measured
    quasi-condensate fractions \cite{Ries2015PRL}. Adapted from
    Ref.~\cite{Wang2020NJP}.  (b) Overlay of our $T_\text{BKT}$ on top
    of a collection of experimental data
    \cite{Ries2015PRL,Lennart2021}, QMC \cite{He2022PRL} and various
    theoretical results
    \cite{Petrov2003PRA,Bighin2016PRB,Bauer2014PRL,Mulkerin2017PRA}.
    The black solid and dashed lines represent $T_\text{BKT}$
    calculated using our pairing fluctuation theory with and without
    the particle-hole channel contributions, respectively. Here the
    particle-hole channel effect was calculated with level 1 averaging
    of $\langle \chi_\text{ph}^{}\rangle$.  Adapted from
    Ref.~\cite{He2022PRL}.  }
    \label{fig:comp}
\end{figure}

Finally, in Fig.~\ref{fig:comp}, we compare our theoretical
$T_\text{BKT}$ with available experimental data \cite{Ries2015PRL} and
QMC results, as well as other theories where appropriate, on the BKT
transition in 2D Fermi gases, as a function of
$\ln(k_\text{F}a_\text{2D})$ throughout the BCS-BEC crossover.  Figure
\ref{fig:comp}(a) presents the measured quasi-condensate fractions
$N_q/N$ \cite{Ries2015PRL}, overlaid on top of which are the theory
curves calculated using our pairing fluctuation theory without (black
solid) and with (light cyan solid curve) the particle-hole
contributions (level 1). The black solid line was previously presented
in Ref.~\cite{Wang2020NJP}, calculated with a fixed ${\cal
  D}_\text{B}=4.9$. The experimental data are not dense enough to
enable a successful detection of the minimum in $T_\text{BKT}$ near
$\ln(k_\text{F}a_\text{2D}) = 0$, nevertheless, it does seem to
suggest there exists a minimum in this neighborhood in the contour of
the quasi-condensate fraction. Thus, our theory is consistent with the
data, including the presence of the minimum. Our present curve does align
better with the slightly decreasing $T_\text{BKT}$ and contour of the
quasi-condensate fraction toward BEC. In Fig.~\ref{fig:comp}(b), we
compare our  $T_\text{BKT}$ with a collection of
other theories
\cite{Petrov2003PRA,Bighin2016PRB,Bauer2014PRL,Mulkerin2017PRA}, and
results obtained from QMC using a 2D lattice model \cite{He2022PRL}.
Also plotted are the experimental results from
Refs.~\cite{Ries2015PRL} and \cite{Lennart2021}.
In the BEC regime, where particle-hole contributions become
negligible, our results both with (black solid) and without (black
dashed line) are in quantitative agreement with both experiments, as
well as that of Petrov et al \cite{Petrov2003PRA}. Here we choose to
show only the result with level 1 average of the particle-hole
susceptibility, giving the relatively small quantitative difference in
$T_\text{BKT}$ between levels 1 and 2. There is an expected small
difference between Petrov et al \cite{Petrov2003PRA} and our present
result based on experimental ${\cal D}_\text{B}$, due to the
quantitative difference between the pair size used in
Ref.~\cite{Petrov2003PRA} and the scattering length $a_\text{2D}$. In
the BCS regime, the curves of Mulkerin et al \cite{Mulkerin2017PRA}
and of the BCS mean-field treatment are close to our curve without
particle-hole, suggesting that particle-hole contributions are not
included in Mulkerin et al \cite{Mulkerin2017PRA}. Our particle-hole
corrected result (black solid curve) is in good agreement with the QMC
result extrapolated to the infinite $L$ limit, except that QMC does
not show a minimum in the unitary regime \cite{He2022PRL}. For
$\ln (k_\text{F}a_\text{2D}) > 2$, the QMC for infinite lattice size
$L$ does not have data points, while the results of Petrov et al
\cite{Petrov2003PRA} starts to merge with our (black solid)
particle-hole corrected curve. In comparison, Bauer et al
\cite{Bauer2014PRL} has a higher $T_\text{BKT}$ in this regime; their
curve intersects with the curves of our particle-hole corrected
result, Petrov et al and QMC (infinite $L$).  We also notice a large
difference between the finite ($L=45$) and infinite lattice QMC
results. It is unusual that the result of Bighin et al
\cite{Bighin2016PRB} is far above the BCS mean-field value in the BCS
limit \footnote{We note that, in Ref. \cite{Bighin2016PRB}, the
  expressions for the normal fluid densities $n_{n,f}$ and $n_{n,b}$
  in their Eqs. (15-16) were wrong by a dimensionful factor
  $1/m$.}. They also ignored the particle-hole contributions.  It
should be noted that the experimental data in the BCS regime are
significantly above all theory curves, except the mean-field result;
no sign of decrease toward the BCS regime can be discerned. This
suggests that some other factors, e.g., nonequilibrium, finite size
effect, need to be considered. Therefore, we shall refrain from
comparing with these experimental data in the BCS regime.
In short, our results with the particle-hole contributions are
consistent with experiment in the unitary and BEC regimes and in good
agreement with the QMC results \cite{He2022PRL} in the unitary and BCS
regimes. This calls for more data from future experiment.

A few remarks are in order. (i) In the BCS limit, the bosonic
fluctuation contributions are relatively weak. The mean-field
superfluid density $n_s/m$ has a strong temperature dependence and
decreases sharply toward zero at the mean-field transition temperature
$T_c^\text{MF}$. Any fluctuation effect would only suppress
$n_s/m$. Therefore, it serves to provide an upper bound for
$T_\text{BKT}$. The sharp $T$ dependence of $n_s/m$ (and
$n_\text{B}/M_\text{B}$) near $T_c^\text{MF}$ implies that one has
$T_\text{BKT}\lesssim T_c^\text{MF}$, insensitive to the actual
condition used.
(ii) It is well known that, in the continuum, one has
$n_s(T=0)/m = n/m$ due to the Galilean translational invariance. For
this reason, Shi et al. \cite{Shi2024} presented an
interaction-independent upper bound $T_F/8$ for $T_\text{BKT}$.  Thus
it is hard to imagine that $n_s(T=T_\text{BKT})/m$ would properly
reflect the effect of inter-fermion interactions. In this sense, we
suspect that the NKC cannot properly capture the interaction effect in
the crossover regime of a fermionic superfluid system. (iii)
Furthermore, in an NKC-based approach
\cite{Bighin2016PRB,Mulkerin2017PRA}, it is often assumed $M_B=2m$
throughout the BCS-BEC crossover, ignoring possible pair mass
renormalization. (iv) Therefore, the non-monotonic behavior of
$T_\text{BKT}$ we find in Fig.~\ref{fig:Tc}(a) is unlikely to be
present in an NKC-based calculation.

Finally, we note that the particle-hole susceptibility in Eq.~(\ref{eq:chiph}) is density independent in 2D. This suggests that the particle-hole contributions in the weak coupling limit would have already been automatically included, if the 2D scattering length $a_\text{2D}$ were measured directly through experiment, e.g., by measuring the two-body binding energy in the dilute limit. Instead,  $a_\text{2D}$ is usually calculated from the 3D scattering length as in a deeply confined pancake-shaped trap \cite{petrov2001PRA}, thus the effect of particle-hole fluctuations should be seriously taken into account when comparing experiment and theory. In Fig.~\ref{fig:comp}, the same definition of  $a_\text{2D}$ was used in both QMC and our present work. This makes it possible to have a good agreement in the BCS regime between QMC and our result with particle-hole fluctuations included. Conversely, one can experimentally determine the particle-hole susceptibility in the BCS limit by measuring the 2D scattering length and comparing with that calculated from $a_\text{3D}$, and obtain
$$\langle \chi_\text{ph}\rangle = \frac{m}{2\pi}\ln \frac{a^\text{exp}_\text{2D}}{a^{}_\text{2D}}\,,$$
where $a^\text{exp}_\text{2D}$ denotes the experimentally measured 2D
scattering length. Nevertheless, away from the weak coupling limit, a
nontrivial density dependence should emerge as both a nonzero pairing gap
develops and the ratio $\mu/E_\text{F}$ starts to decrease with
increasing pairing strength at a finite density.

\section{Conclusions}

In summary, we have studied the impact of the particle-hole channel on
the BKT physics in Fermi gases within the context of BCS-BEC
crossover. The entire particle-hole $T$ matrix is included to
renormalize the interaction that appears in the particle-particle $T$
matrix. Both $T$ matrices contain self-consistently includes the
self-energy feedback. This leads to a non-uniform shift of the inverse
interaction strength by the particle-hole susceptibility, which
evaluated as an angular average at two different levels, revealing
important physical consequences in the crossover and BCS regimes.  The
particle-hole channel provides a screening of the pairing interaction
and thus shifts curves of the BKT transition temperature and the
pairing gap towards the BEC regime as a function of
$\ln (k^{}_Fa^{}_\text{2D})$.  Additionally, a comparison shows that
the particle-hole corrected BKT transition temperature exhibits
a better agreement with the experimental data and QMC results. Future
experiments with more elaborate measurements are called for to resolve
various discrepancies.

Finally, it should be mentioned that, in addition to the simple
averaged particle-hole susceptibility, there are a series of higher
order corrections, including a higher order $T$-matrix
\cite{Chen2016SR} and vertex corrections, which may contribute
significant modifications to the present results. In addition, the BKT
criterion \cite{Wu2015PRL,Wang2020NJP} is merely based on numerical
simulations \cite{Prokofev2001PRL} to provide the critical phase
space density. Despite the experimental support \cite{Murthy2015PRL},
a rigorous analytical derivation of such a criterion would be highly
desirable.

\section{Acknowledgments}
This work was supported by the Innovation Program 
for Quantum Science and Technology (Grant No. 2021ZD0301904). 

\appendix

\section{Slope discontinuity in $ \chi_{\text{ph}}$ across $\mu=0$ at zero $T$}
\label{sec:AppA}

The expression for the particle-hole susceptibility $\chi_{\text{ph}}(Q')$ is given by 
\begin{eqnarray*}
  &&\chi_{\text{ph}}(Q') =\\
  &&\sum_{\mathbf{k}}\left[\frac{f(E_{\mathbf{k}})-f(\xi_{\mathbf{k}-{\mathbf{q}'}})}{E_{\mathbf{k}}-\xi_{\mathbf{k}-{\mathbf{q}'}}-i \Omega'_n} u_{\mathbf{k}}^2
  -\frac{1-f(E_{\mathbf{k}})-f(\xi_{\mathbf{k}-{\mathbf{q}'}})}{E_{\mathbf{k}}+\xi_{\mathbf{k}-{\mathbf{q}'}}+i \Omega'_n} v_{\mathbf{k}}^2\right].
\end{eqnarray*}
Upon analytical continuation, $\mathrm{i}\Omega'_n \rightarrow \Omega' + \mathrm{i}0^+$, 
we separate the retarded $\chi^R_{\text{ph}}(\Omega',{\mathbf{q}'})$ into real and imaginary parts, 
$\chi^R_{\text{ph}}(\Omega',{\mathbf{q}'})=\chi^\prime_{\text{ph}}(\Omega',{\mathbf{q}'})+\mathrm{i}\chi^{\prime\prime}_{\text{ph}}(\Omega',{\mathbf{q}'})$.
Furthermore, we set $\mathrm{i}\Omega'_n = 0$, which leads to $\chi^{\prime\prime}_{\text{ph}}(0,{\mathbf{q}'})=0$, 
and the real part is expressed as
\begin{eqnarray*}
  &&\chi^\prime_{\text{ph}}(0,{\mathbf{q}'}) = \\
  &&\sum_{\mathbf{k}} \left[\frac{f(E_{\mathbf{k}})-f(\xi_{\mathbf{k}-{\mathbf{q}'}})}{ E_{\mathbf{k}}-\xi_{\mathbf{k}-{\mathbf{q}'}}}u^2_{\mathbf{k}} - 
    \frac{1 -f( E_{\mathbf{k}})-f(\xi_{\mathbf{k}-{\mathbf{q}'}})}{ E_{\mathbf{k}}+\xi_{\mathbf{k}-{\mathbf{q}'}}}v^2_{\mathbf{k}}\right].
\end{eqnarray*}
At $T=0$, we have $f(x) = 1-\Theta(x)$. 
Thus $\chi^\prime_{\text{ph}}(0,{\mathbf{q}'})$ is given by 
\begin{equation*}
    \chi^\prime_{\text{ph}}(0,{\mathbf{q}'}) = 
    \sum_{\mathbf{k}} \left[\frac{\Theta(\xi_{\mathbf{k}-{\mathbf{q}'}})-1}{ E_{\mathbf{k}}-\xi_{\mathbf{k}-{\mathbf{q}'}}}u^2_{\mathbf{k}} - 
    \frac{\Theta(\xi_{\mathbf{k}-{\mathbf{q}'}})}{ E_{\mathbf{k}}+\xi_{\mathbf{k}-{\mathbf{q}'}}}v^2_{\mathbf{k}}\right].
\end{equation*}
Then the derivative of $\chi^\prime_{\text{ph}}(0,{\mathbf{q}'})$ with respect to $\mu$ is given by 
\begin{eqnarray*}
  &&\frac{\partial}{\partial \mu} \chi^\prime_{\text{ph}}(0,{\mathbf{q}'}) = 
  -\sum_{\mathbf{k}} \left[\frac{1+ \delta(\xi_{\mathbf{k}-{\mathbf{q}'}})}{ E_{\mathbf{k}}+\xi_{\mathbf{k}-{\mathbf{q}'}}}v^2_{\mathbf{k}} -  \frac{\delta(\xi_{\mathbf{k}-{\mathbf{q}'}})}{ E_{\mathbf{k}}-\xi_{\mathbf{k}-{\mathbf{q}'}}}u^2_{\mathbf{k}}\right] \\
&& = 
\left\{
\begin{aligned}
	& -\sum_{\mathbf{k}} \frac{v^2_{\mathbf{k}}}{ E_{\mathbf{k}}+\xi_{\mathbf{k}-{\mathbf{q}'}}} \,, & \mu < 0\,, \\
	& -\sum_{\mathbf{k}} \frac{v^2_{\mathbf{k}}}{ E_{\mathbf{k}}+\xi_{\mathbf{k}-{\mathbf{q}'}}} + \sum_{\phi \in A} \frac{\xi_{\mathbf{k}'}}{ E_{\mathbf{k}'}^2}\,,
	& \mu\geq 0\,,
\end{aligned}
\right.
\end{eqnarray*}
where $A = \left\{\phi: p^2(\cos(\phi)^2-1)/2+\mu \right\}$ with $\mathbf{k}' = p\cos(\phi) \pm \sqrt{p^2(\cos(\phi)^2-1)/2+\mu}$ given by $\xi_{\mathbf{k}-{\mathbf{q}'}} = 0$. 
Thus, an additional term in the derivative of $\chi^\prime_{\text{ph}}(0,{\mathbf{q}'})$ 
emerges when the system changes from $\mu<0$ to $\mu \ge 0$, leading to a slope discontinuity of $\chi_{\text{ph}}$.


\section{Determining the critical phase space density ${\cal D}_B$}
\label{App:D_B}

\begin{figure}
   \centerline{\includegraphics[clip,width=3.4in]{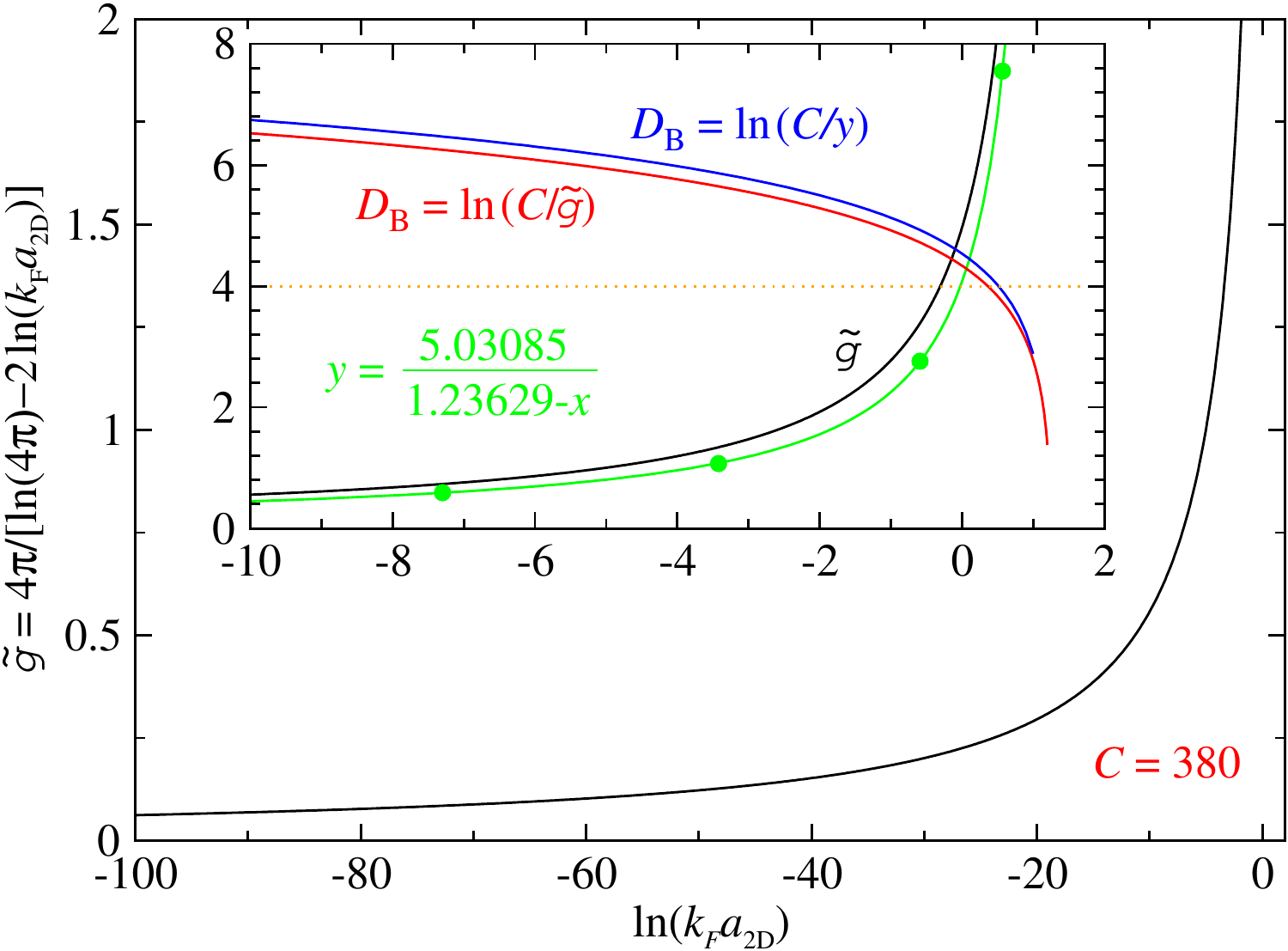}}
  \caption{$\tilde{g}$ as a function of $\ln(k_F a_\text{2D}) $ as
    given by the QMC result of Prokof'ev et al \cite{Prokofev2001PRL} (black solid
    line). Shown in the inset is the corresponding ${\cal D}_\text{B}$ (red
    line), along with the experimental data (green dots) and fit
    (green line) for $\tilde{g}$ from Ref. \cite{Murthy2015PRL}. The fit from
    experimental data yields a slightly larger ${\cal D}_\text{B}$ (blue
    line).}
 \label{fig:g}
\end{figure}

Estimates of $\mathcal{D}_{\text{B}}(T_\text{BKT})$ for fermionic
superfluids are provided in Refs.~\cite{Ries2015PRL,Murthy2015PRL},
with values ranging from approximately 4.9 to 6.45.  In comparison,
the analogous systems in atomic Bose gases typically hover around 8
\cite{Tung2010PRL}.  The lowest value,
$\mathcal{D}_{\text{B}}(T_\text{BKT}) = 4.9$, which is the closest to
the factor of 4 in the usual Nelson-Kosterlitz BKT relation, was reported to offer the
best fit for the experimental results on Fermi gases
\cite{Murthy2015PRL}. In Ref.~\cite{Wang2020NJP}, 
$\mathcal{D}_{\text{B}}(T_\text{BKT})$ was taken at fixed value of 4.9. Here we consider a broader range of interaction and thus include the loglog dependence, albeit very weak, of $\mathcal{D}_{\text{B}}$ on $a_\text{2D}$, and take the data from experiment. To this end, we take the data of $\tilde{g}$ from the Supplemental Materials of Ref.~\cite{Murthy2015PRL}, and plot in Fig.~\ref{fig:g}. The data fits perfectly to Eq.~(\ref{eq:g}) in the text. Also plotted in the figure is the function with the identification of potential radius \cite{Prokofev2001PRL} or pair size \cite{Petrov2003PRA} with $a_\text{2D}$. The small difference between these two curves reveals that  $a_\text{2D}$ differs from either  potential radius or pair size.

\bibliography{2DBKTph.bib}

\end{document}